%% file: green.tex
\documentclass[graybox]{svmult}

\usepackage{mathptmx}       % selects Times Roman as basic font
\usepackage{helvet}         % selects Helvetica as sans-serif font
\usepackage{courier}        % selects Courier as typewriter font
\usepackage{type1cm}        % activate if the above 3 fonts are
\usepackage{makeidx}         % allows index generation
\usepackage{graphicx}        % standard LaTeX graphics tool
\usepackage{multicol}        % used for the two-column index
\usepackage[bottom]{footmisc}% places footnotes at page bottom

\makeindex             % used for the subject index
\begin{document}

\title*{Primordial Black Holes: sirens of the early Universe}
\author{Anne M. Green}
\institute{Anne M. Green \at School of Physics and Astronomy, University of Nottingham, University Park, Nottingham, NG7 2RD, UK \email{anne.green@nottingham.ac.uk}}
%\and Name of Second Author \at Name, Address of Institute \email{name@email.address}}
\maketitle

\abstract{Primordial Black Holes (PBHs) are, typically light, black holes which can form in the early Universe. There are a number of formation mechanisms, including the collapse of large density perturbations, cosmic string loops and bubble collisions. The number of PBHs formed is tightly constrained by the consequences of their evaporation and their lensing and dynamical effects. Therefore PBHs are a powerful probe of the physics of the early Universe, in particular models of inflation. They are also a potential cold dark matter candidate.}

\section{Introduction}

Primordial Black Holes (PBHs) are black holes which may form in the early Universe~\cite{Zeldovich:1967,Hawking:1971ei}. There are various formation mechanisms: the collapse of large density fluctuations (Sec.~\ref{sec-form-dens}), cosmic string loops~\cite{Hawking:1987bn} (Sec.~\ref{sec-form-cs}) or bubble collisions~\cite{Crawford:1982,Hawking:1982}
(Sec.~\ref{sec-form-bubble}). In most cases the PBH mass, $M_{\rm PBH}$, is roughly equal to the horizon mass, $M_{\rm H}$, at the formation epoch (e.g. Ref.~\cite{Carr2005}):
\begin{equation}
M_{\rm PBH} \sim M_{\rm H} \sim \frac{c^3 t}{G} \sim 10^{15} \left( \frac{t}{10^{-23} \, {\rm s} } \right) \, {\rm g} \,.
\end{equation}
For instance PBHs formed at the QCD phase transition at $t \sim 10^{-6} \, {\rm s}$ would have mass of order a solar mass, $M_{\rm PBH} \sim M_{\odot}= 2 \times 10^{30} \, {\rm kg}$.

As famously realised by Hawking~\cite{Hawking:1974rv}, PBHs radiate thermally and hence evaporate on a timescale, $\tau(M_{\rm PBH})$, (e.g. Ref.~\cite{Carr2005}):
\begin{equation}
\tau(M_{\rm PBH}) \sim \frac{\hbar c^4}{G^2 M_{\rm PBH}^3} \sim 10^{10} \left( \frac{M_{\rm PBH}}{10^{15} \, {\rm g}} \right)^3 \, {\rm Gyr} \,.
\end{equation}
PBHs with $M_{\rm PBH} \sim 10^{15} \, {\rm g}$ will be evaporating today and their abundance is constrained by the flux of $\gamma$-rays~\cite{Page:1976wx} (Sec.~\ref{sec-abund-evap-gamma}). Lighter PBHs evaporated in the past and are constrained by the effects of their evaporation products on Big Bang Nucleosynthesis~\cite{Vainer:1977,Zeldovich:1977} (Sec.~\ref{sec-abund-evap-nucleo}) and the present day density of any stable relic particles~\cite{MacGibbon:1987my}  (Sec.~\ref{sec-abund-evap-relic}). Heavier PBHs are stable and their abundance is limited by their lensing (Sec.~\ref{sec-abund-lens}) and dynamical~\cite{Carr:1997cn} (Sec.~\ref{sec-abund-dyn}) effects and also their effects on various other astrophysical processes and objects (Sec.~\ref{sec-abund-other}).
Since PBHs are matter, the fraction of the total energy density in the form of PBHs increases proportional to the scale factor, $a$, during radiation domination. Therefore the constraints on the fraction of the initial energy density in the form of PBHs, $\beta(M_{\rm PBH})= \rho_{\rm PBH}/\rho_{\rm tot}$, are very tight, lying in the range $10^{-5} - 10^{-30}$. 

Cosmological inflation, a period of accelerated expansion in the early Universe, may have generated the primordial fluctuations from which galaxies and large scale structure form (see e.g. Ref.~\cite{Lyth:2009zz}). The constraints on the initial fraction of the energy density in the form of PBHs can be translated into limits on the primordial power spectrum of density perturbations on small scales, and can therefore be used to constrain models of inflation~\cite{Carr:1994ar} (Sec.~\ref{sec-inf}). 

Finally there is extensive astronomical and cosmological evidence that the majority of the matter in the universe is in the form of non-baryonic cold dark matter (CDM) (see e.g. Ref.~\cite{Bertone:2004pz}).  Since PBHs form before nucleosynthesis they are non-baryonic and therefore a candidate for the CDM (Sec.~\ref{sec-dm}).

\section{PBH formation mechanisms}
\label{sec-form}
For a PBH to form a large over-density is required. In this section we discuss several ways of achieving this:  large density fluctuations~\cite{Carr:1974nx} (Sec.~\ref{sec-form-dens}), cosmic string loops~\cite{Hawking:1987bn} (Sec.~\ref{sec-form-cs}) and bubble collisions~\cite{Crawford:1982,Hawking:1982}.

\subsection{Large density fluctuations}
\label{sec-form-dens}
During radiation domination, if a density fluctuation is sufficiently large, then gravity overcomes pressure forces and the fluctuation collapses to form a PBH shortly after it enters the horizon~\cite{Carr:1974nx}. We review the original calculations of this process in Sec.~\ref{sec-form-dens-simp} and then discuss refinements in Sec.~\ref{sec-form-dens-refine}.

\subsubsection{Original calculations}
\label{sec-form-dens-simp}
The early pioneering calculations by Carr and Hawking~\cite{Carr:1974nx} assumed that the over-dense region from which a PBH formed was spherically symmetric~\footnote{PBHs form from the rare large density fluctuations for which this assumption is justified~\cite{Doroshkevich,Bardeen:1985tr}. } and part of a closed Friedmann universe. In this case, for gravity to overcome pressure at maximum expansion the region must be larger than the Jeans length, which is $\sqrt{w}$ times the horizon length ($w$ is the equation of state parameter, $p= w \rho$, and $w=1/3$ for radiation domination). This leads to a requirement that the density contrast, $\delta \equiv \delta \rho/\rho$, at horizon crossing must exceed a critical value $\delta_{\rm c} \approx w$.  It was thought at this time that if the fluctuation was larger than the horizon length, which corresponds to $\delta > 1$, then it would instead form a separate closed universe.

The PBHs formed would have mass of order the horizon mass, $M_{\rm H}$, at the time they form: $M_{\rm PBH} = w^{3/2} M_{\rm H}$~\cite{Carr:1975qj}. If the fluctuations are scale invariant, so that PBHs form on all scales, then the PBHs will have an extended mass function~\cite{Carr:1975qj}: ${\rm d} n/{\rm d} M_{\rm PBH} \propto M_{\rm PBH}^{-5/2}$. However, as we will see in Sec.~\ref{sec-inf-translate}, it is now known that for a scale-invariant power spectrum, normalised to observations on cosmological scales, the number of PBHs formed is completely negligible~\cite{Carr:1993aq}.

The criteria for PBH formation in a matter dominated universe with $w=0$ are somewhat different. In this case, because 
the pressure is zero, it is possible for PBHs to form well within the horizon. However for this to happen the perturbation must be sufficiently spherically symmetric~\cite{Khlopov,Polnarev}.

\subsubsection{Refinements}
\label{sec-form-dens-refine}

Early numerical simulations of PBH formation~\cite{Nadezhin} roughly confirmed the earlier analytic calculations~\cite{Carr:1974nx,Carr:1975qj}. More recently it has been realised that, as a consequence of near critical gravitational collapse~\cite{Choptuik:1992jv}, the PBH mass depends on the size of the fluctuation from which it formed~\cite{Niemeyer:1997mt,Niemeyer:1999ak}:
\begin{equation}
M_{\rm PBH} = \kappa M_{\rm H} (\delta-\delta_{\rm c})^{\gamma} \,,
\end{equation}
where $\gamma$ and $\kappa$ are constants of order unity which depend on the shape of the perturbation and the background equation of state~\cite{Niemeyer:1997mt,Niemeyer:1999ak,Musco:2004ak}. This power law scaling of the PBH mass has been found to hold down to $(\delta-\delta_{\rm c}) \sim 10^{-10}$ ~\cite{Musco:2008hv,Musco:2012au}. Note, however, that the majority of the PBHs formed 
have masses within an order of magnitude or so of $M_{\rm H}$~\cite{Niemeyer:1999ak}. For scale-dependent power spectra which produce an interesting PBH abundance it can be assumed that all PBHs form at a single epoch~\cite{Green:1999xm}. For a power spectrum which increases monotonically with decreasing scale, PBH formation occurs at the smallest scale, while if the power spectrum has a feature, PBH formation occurs on the scale on which the perturbations are largest. The spread in the mass function due to critical collapse can have an important effect on the constraints on the PBH abundance which are sensitive to the PBH mass function, 
for instance the constraint from the flux of gamma-rays~\cite{Yokoyama:1998xd,Kim:1999iv}.

There has been a lot of interest in the exact value of the threshold for PBH formation, $\delta_{\rm c}$, since, as we will see in Sec.~\ref{sec-inf-translate}, the number of PBHs formed depends exponentially on $\delta_{\rm c}$. Ref.~\cite{Shibata:1999zs} found that the peak value of the metric perturbation was a good indicator of PBH formation, and Ref.~\cite{Green:2004wb} found, using peaks theory, that the threshold values from Ref.~\cite{Shibata:1999zs}  were equivalent to density thresholds in the range $\delta_{\rm c} \approx 0.3-0.5$.  Ref.~\cite{Musco:2012au}, which used appropriate initial conditions (super-horizon perturbations which only contain a growing mode) for their simulations, found for radiation domination $\delta_{\rm c} \approx 0.45$. This is in good agreement with  recent analytic calculations, once gauge issues are taken into account~\cite{Harada:2013epa}.

Whether or not a fluctuation collapses to form a PBH depends on its shape as well as its amplitude~\cite{Polnarev:2006aa,Polnarev:2012bi}. Ref.~\cite{Nakama:2013ica} explored fluctuations with a range of shapes and found that the key parameters are the average value of the central overdensity and the width of the overdensity (for broad overdensities the threshold average density is reduced).

 There has also been development regarding the fate of perturbations with $\delta \sim 1$. Ref.~\cite{Kopp:2010sh} showed that they do not form a separate closed universe, and the upper limit was in fact a consequence of the gauge choice.

\subsection{Cosmic string loops}
\label{sec-form-cs}
Cosmic string are one-dimensional topological defects which may form during phase transitions in the very early Universe (see e.g. Ref.~\cite{Vilenkin}).
As a cosmic string network evolves, long strings self-intersect and form cosmic string loops. There is a small probability that an oscillating cosmic string loop will be in a configuration where all of its dimensions are less than its Schwarzschild radius, and hence it will collapse to form a PBH with mass roughly equal to the horizon mass~\cite{Hawking:1987bn,Polnarev:1988dh,Garriga:1993gj,Caldwell:1995fu,Wichoski:1998kh}.
The number of PBHs formed depends on the mass per unit length of the strings, $\mu$, which is related to the symmetry breaking scale. Cosmic string loops can collapse to form PBHs at any point during radiation domination, therefore the resulting PBHs have an extended mass function ${\rm d} n/{\rm d} M_{\rm PBH} \propto M_{\rm PBH}^{-5/2}$. The constraints on the number of PBHs formed (from the gamma-rays and cosmic-rays produced when they evaporate, see Sec.~\ref{sec-abund-evap}) place a limit on the cosmic string mass per unit length $G \mu /c^2 < 10^{-6}$~\cite{Wichoski:1998kh}, comparable to the constraints from the effect of strings on the Cosmic Microwave Background radiation~\cite{Ade:2013xla}.

\subsection{Bubble collisions}
\label{sec-form-bubble}

First order phase transitions occur through the formation of bubbles of the new phase, which then expand and collide.
PBHs, with mass of order the horizon mass, can form as a result of these bubble collisions~\cite{Crawford:1982,Hawking:1982,La:1989st}. However forming a cosmologically interesting abundance of PBHs requires fine tuning of the bubble formation rate, so that the bubbles collide but the phase transition doesn't occur instantaneously.

\section{PBH abundance constraints}
\label{sec-abund}
PBH abundance constraints are usually quoted in terms of the fraction of the energy density in the form of PBHs at the time they form
\begin{equation}
\beta(M_{\rm PBH}) \equiv \frac{\rho_{\rm PBH}}{\rho_{{\rm tot}}} \,.
\end{equation}
The PBH density evolves as $\rho_{\rm PBH} \propto a^{-3}$, where $a$ is the scale factor, while the radiation density varies as  $\rho_{\rm rad} \propto a^{-4}$. Therefore during radiation domination the fraction of the total energy density which is in the form of PBHs grows proportional to $a$. So even if the fraction of the energy density of the Universe in PBHs is initially small it can grow to be significant at late times.

The lensing (Sec.~\ref{sec-abund-lens}) and dynamical (Sec.~\ref{sec-abund-dyn}) constraints limit the fraction of the Milky Way (MW) halo in the form of compact objects, $ f(M_{\rm CO})$. Assuming the density of other compact object is negligible and the MW's DM composition is the same as the Universe as a whole (which is a reasonable assumption) then $f(M_{\rm CO})=\Omega_{\rm PBH}/\Omega_{\rm CDM}$,
where $\Omega_{\rm PBH}$ and $\Omega_{\rm CDM}$ are the fraction of the critical density (for which the geometry of the Universe is flat) in the form of PBHs and CDM respectively. It is related to the initial PBH mass fraction, $\beta(M_{\rm PBH})$, via (see e.g. Ref.~\cite{Carr:2009jm} for a more accurate expression)
\begin{equation}
f(M_{\rm CO}) \approx \left( \frac{\beta(M_{\rm PBH})}{10^{-8}} \right) \left( \frac{M_{\rm PBH}}{M_\odot} \right)^{-1/2} \,.
\end{equation}

Most of the constraints that we discuss below effectively apply to the integral of the PBH mass function over the range of applicability. This range is usually significantly larger than the expected width of the PBH mass function (see Sec.~\ref{sec-form-dens}), and therefore the constraints are not sensitive to the precise form of the mass function. However the constraints from cosmic-rays and gamma-rays produced by recently and currently evaporating PBHs are an exception to this (see e.g. Ref.~\cite{Yokoyama:1998xd,Kim:1999iv}). 

For conciseness we only give order of magnitude values of the constraints and their range of validity,  see Ref.~\cite{Carr:2009jm} for the precise mass dependence of the constraints on the halo fraction in compact objects, $f(M_{\rm CO})$, and the initial mass fraction of PBHs, $\beta(M_{\rm PBH})$.

\subsection{Evaporation}
\label{sec-abund-evap}

The current picture of PBH evaporation~\cite{MacGibbon:1990zk} is that they directly emit all particles which appear elementary at the energy scale of the PBH and have rest mass less than the black hole temperature. Therefore if the black hole temperature exceeds the QCD confinement scale, quark and gluon jets are emitted directly. The quark and gluon jets then fragment and decay producing astrophysically stable particles: photons, neutrinos, electrons, protons and their anti-particles. It has been argued that QED or QCD interactions could lead to the formation of an optically thick photosphere~\cite{Heckler:1995qq,Heckler:1997jv}, however the emitted particles do not interact enough for this to occur~\cite{MacGibbon:2007yq}.

There are also limits from photons distorting the spectrum of the Cosmic Microwave Background radiation~\cite{Zeldovich:1977,Carr:2009jm}, the Super-Kamionkande limit on the flux of relic anti-neutrinos~\cite{Carr:2009jm} and extragalactic anti-protons~\cite{Carr:2009jm}. However these constraints are weaker than the nucleosynthesis and extragalactic gamma-ray background constraints~\cite{Carr:2009jm}, so we do not discuss them in detail here.

See Ref.~\cite{Carr:2009jm} for detailed discussion of the evaporation constraints.

\subsubsection{Entropy}
The photons emitted by PBHs with $M_{\rm PBH}<10^{9} \, {\rm g}$ will thermalize and contribute to the baryon to photon ratio. The requirement that this ratio must not exceed the observed value of $\sim 10^{9}$ leads to a relatively weak constraint, $\beta(M_{\rm PBH}) < 10^{-5} (M_{\rm PBH}/10^{9} \, {\rm g})^{-1}$ for $10^{6} \, {\rm g} <M_{\rm PBH} < 10^{9} \, {\rm g}$~\cite{Zeldovich:1977}.

\subsubsection{Relic particles}
\label{sec-abund-evap-relic}
It has been argued that black hole evaporation could leave a stable Planck mass relic~\cite{MacGibbon:1987my}, in which case the present day density of relics must not exceed the upper limit on the present day CDM density. This leads to a constraint of order $\beta(M_{\rm PBH}) < (M_{\rm PBH}/10^{15} \, {\rm g})^{3/2}$ for $M_{\rm PBH}<10^{15} \, {\rm g}$.

In many extensions of the standard model there are stable or long lived massive (${\cal O}(100\, {\rm GeV/c}^2)$) particles. PBHs with mass $M_{\rm PBH} < 10^{11} \, {\rm g}$ can emit these particles and their abundance is consequently limited to be, roughly, \\
$\beta(M_{\rm PBH}) < 10^{-18} (M_{\rm PBH}/10^{11} \, {\rm g})^{-1/2}$ by the present day abundance of stable massive particles~\cite{Green:1999yh} and the decay of long-lived particles~\cite{Lemoine:2000sq,Khlopov:2004tn}.

The limits from Planck mass relics and (quasi-) stable massive particles are weaker than those from nucleosynthesis, and also depend on the uncertain details of beyond the standard model physics. However they are the only potential constraints on PBHs with $M_{\rm PBH} < 10^{6} \, {\rm g}$ and significantly tighter than the entropy constraints for $10^{6} \, {\rm g} <M_{\rm PBH} < 10^{9} \, {\rm g}$.

\subsubsection{Nucleosynthesis}
\label{sec-abund-evap-nucleo}
Extensive work on the effects of PBH evaporation on the products of Big Bang nuclesynthesis was carried out in the late 1970s~\cite{Zeldovich:1977,Vainer:1977,Miyama:1978mp,Vainer:1978,Naselskii:1978,Lindley1980}. Carr et al.~\cite{Carr:2009jm}, using the results of Refs.~\cite{Kohri:1999ex,Kawasaki:2004yh}, have updated the resulting constraints on the abundance of PBHs, taking into account the latest observational data on the abundances of the light elements and the neutron lifetime, and developments in the understanding of PBH evaporation~\cite{MacGibbon:1990zk}.

PBHs with mass in the range $10^{9} \, {\rm g} <M_{\rm PBH} < 10^{10} \, {\rm g}$ have lifetime $\tau=10^{-2} -10^{-3} \, {\rm s}$ and the mesons and anti-nucleons they emit would increase the neutron/proton ratio and hence the abundance of ${}^4 {\rm He}$. This leads to a constraint which is very roughly $\beta < 10^{-20} (M_{\rm PBH}/10^{10} \, {\rm g})^{-2}$~\cite{Carr:2009jm}.
For masses in the range $10^{10} \, {\rm g} <M_{\rm PBH} < 10^{12} \, {\rm g}$ the lifetime is between $10^{-2} \, {\rm s}$ and $10^{2} \, {\rm s}$  and the high-energy hadrons produced dissociate the light elements, reducing the abundance of  ${}^4 {\rm He}$ and increasing the abundance of the other elements. The tightest constraints are from deuterium
 for $10^{10} \, {\rm g} <M_{\rm PBH} < 5 \times 10^{10} \, {\rm g}$ and from non-thermally produced ${}^6 {\rm Li}$ for 
$ 5 \times 10^{10} \, {\rm g} <M_{\rm PBH} < 10^{12} \, {\rm g}$. For both mass ranges the constraint is, roughly, $\beta(M_{\rm PBH})< 10^{-23}$~\cite{Carr:2009jm}. Finally, for $10^{12} \, {\rm g} <M_{\rm PBH} < 10^{13} \, {\rm g}$ the lifetime is $\tau=10^{7} -10^{12} \, {\rm s}$ and photodissociation is instead important. The most stringent constraint, which is again of order $\beta(M_{\rm PBH})< 10^{-23}$, comes from overproduction of ${}^3 {\rm He}$ or deuterium~\cite{Carr:2009jm}.
For further information, including the detailed mass dependence of the constraints see Ref.~\cite{Carr:2009jm}.

\subsubsection{$\gamma$-rays}
\label{sec-abund-evap-gamma}

PBHs with masses in the range $10^{13} \, {\rm g} < M_{\rm PBH} < 10^{15} \, {\rm g}$ will have evaporated between $z \sim 1000$ and the present day and can contribute to the diffuse extragalactic gamma-ray background~\cite{Page:1976wx,Carr1976,MacGibbon:1991vc}. Their abundance is limited by EGRET data to be roughly $\beta(M_{\rm PBH}) < 10^{-27} (M_{\rm PBH}/10^{15} \, {\rm g})^{-5/2}$~\cite{Carr:2009jm}. Slightly more massive PBHs, that have not evaporated completely by the present day, can also emit a significant flux of gamma-rays, and their abundance is limited to be roughly $\beta(M_{\rm PBH}) < 10^{-26} (M/10^{15} \, {\rm g})^{7/2}$~\cite{Carr:2009jm}. There is also a similar limit on PBHs with $M_{\rm PBH} \sim 10^{15} \, {\rm g}$ that are evaporating today from Galactic gamma-rays~\cite{Carr:2009jm}.

The detailed values of these constraints, in particular those for $M_{\rm PBH} > 10^{15} \, {\rm g}$, depend on the exact shape of the PBH mass function~\cite{Yokoyama:1998xd,Kim:1999iv}. However the values stated are somewhat conservative as known astrophysical backgrounds have not been subtracted~\cite{Carr:2009jm}.

For PBHs with $10^{15} \, {\rm g} < M_{\rm PBH} < 10^{17} \, {\rm g}$ the gamma-ray constraints constrain the fraction of the DM in the form of PBHs to be less than one. In other words, PBHs in this mass range can not make up all of the DM.

\subsection{Lensing}
\label{sec-abund-lens}
If there is a cosmologically significant density of compact objects (COs) then there is a high probability that a distant point source is lensed~\cite{Press:1973}.  For hundred solar mass and lighter lenses the image separation is too small for multiple images to be observed, however other observable effects can occur.

\subsubsection{ Gamma-ray burst femtolensing}
\label{sec-abund-lens-grb}

%For COs with mass in the range $10^{17} \, {\rm g} < M < 10^{26} \, {\rm g}$ the image separation is of order pico or femto-arc-seconds. 

For $10^{17} \, {\rm g} < M_{\rm CO} < 10^{20} \, {\rm g}$ the image separation is of order femto arc-seconds.
However the time delay between images, $10^{-17}-10^{-20} \, {\rm s}$, is approximately equal to the period of a gamma-ray. COs of this mass could therefore be detected by the interference pattern in the energy spectrum of a gamma-ray burst~\cite{Gould92}. Analysis of data from the Gamma-ray Burst Monitor onboard the Fermi satellite finds that for  $10^{17} \, {\rm g} < M_{\rm CO} < 10^{20} \, {\rm g}$, $f(M_{\rm CO})<1$~\cite{Barnacka:2012bm}.

%COs with $10^{18} \, {\rm g} < M < 10^{26} \, {\rm g}$ have Einstein radii of order an astronomical unit, and hence detectors separated  by more than this distance would not both observe a particular gamma-ray burst to be lensed ***REWRITE***~\cite{Nemiroff}. 
%Data from the BATSE and Ulysses detectors place a weak limit, larger than the measured DM density, on COs with $10^{21} \, {\rm g} < M < 10^{26} \, {\rm g}$ ~\cite{Marani}. 

\subsubsection{Galactic microlensing}
\label{sec-abund-lens-micro}
Microlensing occurs when a CO with mass in the range $10^{24} \, {\rm g} < M_{\rm CO} < 10^{34} \, {\rm g}$ crosses the line of sight to a star. The image separation is too small (of order micro arc-seconds) for multiple images to be resolved, and the lensing
leads to a temporary amplification of the star's flux~\cite{Paczynski:1985jf}. 
The EROS and MACHO surveys of the Large and Small Magellanic Clouds found that for $10^{26} \, {\rm g} < M_{\rm CO} < 10^{34} \, {\rm g}$, $f(M_{\rm CO})<1$, with tighter limits within this mass range~\cite{Alcock:2000kg,Tisserand:2006zx}. Recently the lower limit of the excluded mass range has been lowered to $4 \times 10^{24} \, {\rm g}$ using Kepler data~\cite{Griest:2013aaa}.

\subsubsection{Quasar microlensing}
COs with $10^{30} \, {\rm g} < M_{\rm CO} < 10^{35} \, {\rm g}$ can microlens quasars, amplifying the continuum emission, without affecting the line emission~\cite{Canizares} and  limits on the number of small equivalent width quasars place a constraint on COs in this mass range $f(M_{\rm CO})<1$~\cite{Dalcanton}.

\subsubsection{Radio source millilensing}
Massive COs with $10^{39} \, {\rm g} < M_{\rm CO} < 10^{41} \, {\rm g}$ can millilens radio sources, producing multiple images which can be resolved with Very Long Baseline Interferometry~\cite{Kassiola}. A null search limits COs in this mass range to make up less than $1\%$ of the total energy density of the Universe~\cite{Wilkinson:2001vv}.

\subsection{Dynamical effects}
\label{sec-abund-dyn}

The abundance of massive COs in the MW halo is constrained by their dynamical effects on the constituents of the MW.
These constraints have been studied in detail by Carr and Sakellariadou~\cite{Carr:1997cn}.
Here we briefly summarise the constraints which place the tightest limits on the halo fraction in COs.
There is also a constraint from the tidal disruption of globular clusters (GCs)~\cite{Carr:1997cn}, however this depends sensitively on the mass and radius of the GCs and is weaker than those from dynamical fiction and disk heating~\cite{Carr:2009jm} so we do not discuss it in detail.

\subsubsection{Disruption of wide binaries}
\label{sec-abund-dyn-bin}
Encounters with massive COs can disrupt~\cite{Bahcall85} or change the orbital parameters~\cite{Weinberg87} of wide binary stars. Observations of wide binaries~\cite{Chaname:2003fn,Quinn:2009zg} constrain the halo mass fraction to be $f(M_{\rm CO})<0.4$
for $10^{36} \, {\rm g} < M_{\rm CO} < 10^{41} \, {\rm g}$~\cite{Quinn:2009zg,Carr:2009jm}.

\subsubsection{Dynamical friction}
COs will be dragged into the centre of the MW by the dynamical friction of spheroid stars and the population of COs themselves. Constraints on
the central mass of the MW limit the halo fraction in COs with $10^{37} \, {\rm g} < M_{\rm CO} < 10^{45} \, {\rm g}$. The limit is tightest, $f(M_{\rm CO}) <  5 \times 10^{-5}$, at $M_{\rm CO} \sim 10^{41} \, {\rm g}$~\cite{Carr:1997cn,Carr:2009jm}.

\subsubsection{Disk heating}
\label{sec-abund-dyn-disc}
Massive COs traversing the Galactic disk will heat the disk, increasing the velocity dispersion of the disk stars~\cite{Lacey85}.
This leads to a limit, from the observed stellar velocity dispersions, on the halo fraction in COs with $10^{40} \, {\rm g} < M_{\rm PBH} < 10^{45} \, {\rm g}$ which is tightest, $f(M) < 10^{-3}$, at $M \sim 10^{43} \, {\rm g}$~\cite{Carr:1997cn,Carr:2009jm}.

\subsection{Other astrophysical objects and processes}
\label{sec-abund-other}
There are also constraints on PBHs with $M_{\rm PBH}> 10^{15} \, {\rm g}$ from their effects on  various astrophysical objects and processes.

\subsubsection{Stars}

If a PBH is captured by a neutron star the star will be accreted and destroyed in a short time~\cite{Capela:2013yf}. The existence of neutron stars in globular clusters (GCs) excludes PBHs with $10^{18} \, {\rm g} < M_{\rm PBH} < 10^{24} \, {\rm g}$ comprising all of the DM, if the DM density in GCs is larger than $ \sim 100 \, {\rm GeV} \, {\rm cm}^{-3}$~\cite{Capela:2013yf}. Similarly accretion of PBHs during star formation could exclude PBH DM in the range $10^{16} \, {\rm g} < M_{\rm PBH} < 10^{22} \, {\rm g}$~\cite{Capela:2012jz}.
Note, however, that a high DM density in GC is only expected if (a subset of) GCs are formed in DM mini halos rather than from baryonic processes and there is no direct observational evidence for a large DM density in GCs (see e.g. Ref.~\cite{Brodie:2006sd}). 
Recently Ref.~\cite{Pani:2014rca} has argued that the existence of old neutron stars in the centres of the MW and Large Magellanic Cloud excludes PBHs in the range $10^{17} \, {\rm g} < M_{\rm PBH} < 10^{24} \, {\rm g}$ comprising all of the DM.

\subsubsection{Gravitational waves}
\label{sec-abund-other-gw}
For PBHs formed from the collapse of large density perturbations there is an indirect limit on their abundance from gravitational waves.
Large density perturbations generate second order tensor perturbations and therefore limits on their amplitude constrain the amplitude of the density perturbations and hence the abundance of PBHs formed. Pulsar timing constraints place a tight limit on the present day density parameter of PBHs with $10^{35} \, {\rm g} < M_{\rm PBH} < 10^{37}  \, {\rm g}$: $\Omega_{\rm PBH} h^2 \leq 10^{-5}$ (where $h \approx 0.7$ is the dimensionless Hubble constant)~\cite{Saito:2008jc}.

\subsubsection{Cosmic Microwave Background radiation}

After decoupling massive PBHs can accrete material and the subsequent radiation can affect the thermal history of the Universe~\cite{Carr1981}. X-rays emitted by gas accreted onto PBHs modify the cosmic recombination history, producing measurable effects on the spectrum and anisotropies in the Cosmic Microwave Background radiation, which have been constrained using the FIRAS and WMAP data~\cite{Ricotti:2007au}. The precise limits depend on the accretion model, but are of order $f(M_{\rm PBH})<10^{-2}$ for $10^{34} \, {\rm g}< M_{\rm PBH} < 10^{36} \, {\rm g}$ and $f(M_{\rm PBH})<10^{-5}$ for $10^{36} \, {\rm g}< M_{\rm PBH} < 10^{42} \, {\rm g}$, with weaker constraints outside these mass ranges~\cite{Ricotti:2007au}.

\subsubsection{Large scale structure}
Massive PBHs would affect large scale structure formation due to the Poisson fluctuations in their number density which enhance the DM power spectrum~\cite{Meszaros}. Lyman-alpha forest observations constrain the fraction of the DM in PBHs with $10^{37} \, {\rm g} < M_{\rm PBH} < 10^{43} \, {\rm g}$~\cite{Afshordi:2003zb,Carr:2009jm}. The limit is tightest, $f(M_{\rm PBH}) < 10^{-3}$, at $M \sim 10^{40} \, {\rm g}$~\cite{Carr:2009jm}.

\section{Constraints on the primordial power spectrum and inflation}
\label{sec-inf}

Cosmological inflation is a period of accelerated expansion proposed to have occurred in the very early Universe in order to solve various problems with the standard Hot Big Bang (namely the horizon, flatness and monopole problems). Accelerated expansion requires negative pressure, a requirement which is satisfied by a slowly-rolling scalar field. Inflation also provides a mechanism for the generation of the primordial fluctuations from which galaxies and large scale structure later form. During inflation the wavelengths of quantum fluctuations in the inflaton field become larger than the Hubble radius and a spectrum of super-Horizon curvature perturbations are generated.  After inflation the fluctuations re-enter the horizon and potentially provide the seeds for structure formation.  For an overview see e.g. Ref.~\cite{Lyth:2009zz}.

The amplitude and scale dependence of the primordial fluctuations depend on the inflaton potential. Therefore observational constraints on the power spectrum of the primordial curvature perturbation
\begin{equation}
{\cal P}_{\cal R}(k) \equiv  \frac{k^3}{2 \pi^2}  \langle |{\cal R}_{k} |^2 \rangle \,,
\end{equation}
where ${\cal R}_k$ are the Fourier modes of the curvature perturbation and $k$ is comoving wavenumber, constrain models of inflation.
The power spectrum on cosmological scales, $k \sim 10^{-3}-1 \, {\rm Mpc}^{-1} $, is accurately measured by Cosmic Microwave Background~\cite{Hlozek,Ade} and large scale structure observations~\cite{Bird:2010mp}, and some inflation models are now ruled out~\cite{Ade}.

Cosmological observations probe a fairly limited region of the inflaton potential. As we will see in this section, the PBH constraints on the power spectrum are fairly weak (many order of magnitude larger than the measurements on cosmological scales). However they apply over a very wide range of scales, $k \sim 10^{-2}-10^{23} \, {\rm Mpc}^{-1}$, and therefore constrain a much broader region of the inflaton potential~\cite{Carr:1993aq}, and eliminate or constrain otherwise viable models~\cite{Peiris:2008be,Josan:2010cj}.

\subsection{Translating limits on the PBH abundance into constraints on the primordial power spectrum}
\label{sec-inf-translate}

As we saw in Sec.~\ref{sec-form-dens} a fluctuation on a physical scale $R$ will collapse to form a PBH, with mass $M_{\rm PBH}$ roughly equal to the horizon mass, if the smoothed density contrast at horizon entry, $\delta_{\rm hor}(R)$, exceeds a threshold value $\delta_{\rm c}$ which is slightly less than unity~\footnote{Ref.~\cite{Lyth:2005ze} argues that PBHs can also form on sub-horizon scales which never exit the horizon.}.  Assuming the initial perturbations have a Gaussian distribution then the probability distribution of the smoothed density contrast is
given by (e.g. Ref.~\cite{Lyth:2009zz}):
\begin{equation}
P(\delta_{\rm hor} (R)) = \frac{1}{\sqrt{2 \pi} \sigma_{\rm hor}(R)}
\exp{ \left( - \frac{\delta^2_{\rm hor}(R)}{2 \sigma_{\rm hor}^2(R)} \right)}\,,
\end{equation}
where $\sigma_{\rm hor}(R)$ is the mass variance evaluated when the scale of interest enters the horizon. The mass variance is defined as
\begin{equation}
\label{variance}
\sigma^2(R)=\int_{0}^{\infty} \tilde{W}^2(kR)\mathcal{P}_{\delta}(k, t)\frac{{\rm d} k}{k},
\end{equation}
while $\mathcal{P}_{\delta}(k, t)$ is the power spectrum of the
(unsmoothed) density contrast
\begin{equation}
\mathcal{P}_{\delta}(k, t) \equiv \frac{k^3}{2 \pi^2} \langle
|\delta_{k} |^2 \rangle \,,
\end{equation}
and $\tilde{W}(kR)$ is the Fourier transform of the window function used to smooth the density contrast. See appendix B of Ref.~\cite{Bringmann:2011ut} for the detailed calculation of the relationship between the primordial power spectrum
of the curvature perturbation and the mass variance at horizon crossing.

The initial PBH mass fraction is equal to the fraction of the energy density of the Universe contained in
regions dense enough to form PBHs which is given, as in
Press-Schechter theory~\cite{Press:1973iz}, by
\begin{equation}
\beta(M_{\rm PBH})  = 2 
      \int_{\delta_{\rm c}}^{\infty} P(\delta_{\rm hor}(R)) \,{\rm d} \delta_{\rm hor}(R) \,,
	               \label{presssch}
\end{equation}
where the right hand side is usually multiplied by a factor of 2, so that all of the mass in the Universe is accounted for. The PBH
initial mass fraction is then related to the mass variance by
\begin{eqnarray}	
\beta(M_{\rm PBH})& =&  \frac{2}{\sqrt{2\pi}\sigma_{\rm hor}(R)} 
\int_{\delta_{\rm c}}^{\infty} \exp{\left(- \frac{\delta^2_{\rm hor}(R)}
    {2 \sigma_{\rm hor}^2(R)}\right)} 
  \,{\rm d}\delta_{\rm hor}(R) \,,\nonumber \\
 &=&   {\rm erfc}\left(\frac{\delta_{\rm c}}{
   \sqrt{2}\sigma_{\rm hor}(R)}\right)  \,.     
\label{densitypara}
\end{eqnarray}
Note that if the power spectrum of the density perturbations were exactly scale invariant (so that $\sigma_{\rm hor}$ is independent of $R$) then the abundance of PBHs would be completely negligible,  $\beta(M_{\rm PBH}) \sim \exp{(-10^{8})}$, since on cosmological scales the mass variance is measured to be of order $10^{-5}$~\cite{Carr:1993aq}.

The limits on the PBH abundance, $\beta(M_{\rm PBH})$, can be translated into constraints on the power spectrum of the primordial curvature perturbation by first inverting eq.~(\ref{densitypara}) to find the limits on the mass variance at horizon crossing, $\sigma_{\rm hor}(R)$. Since the mass variance is given by an integral over the primordial power spectrum, it depends not just on the amplitude of the power spectrum on a single scale, but also its shape on neighbouring scales. In practice, however, if the power spectrum is featureless then it is a good approximation to parameterize it {\em locally} as a power-law and the constraints depend only weakly (at the per-cent level) on the slope of the power-law~\cite{Josan:2009qn,Bringmann:2011ut}. The constraints on
the initial abundance of PBHs, which lie in the range $\beta(M_{\rm PBH})< 10^{-30}-10^{-5}$, translate into constraints on the amplitude of the power spectrum of the primordial curvature perturbation in the range ${\cal P}_{\cal R}< 10^{-2}-10^{-1}$~\cite{Bugaev:2008gw, Josan:2009qn}. See Fig. 2 of Ref.~\cite{Josan:2009qn} for a plot of the detailed scale dependence of the constraints on the power spectrum.

The standard calculation as described above assumes that the probability distribution function (pdf) of the perturbations is gaussian. Since PBHs form from the extremely rare large fluctuations in the tail of the distribution, non-gaussianity can have a significant effect on the abundance of PBHs formed~\cite{Bullock:1996at,Ivanov:1997ia}. Refs.~\cite{Byrnes:2012yx,Young:2013oia} jointly constrained the amplitude of the fluctuations and the local non-gaussianity parameters, ($f_{\rm nl}$, $g_{\rm nl}$, etc.)  while Ref.~\cite{Shandera:2012ke}
constrained the amplitude for two different physically motivated ansatzes for the scaling of the dimensionless moments of the density contrast. For some specific models (e.g. the curvaton where the primordial fluctuations arise from fluctuations of a second light scalar field which is subdominant during inflation~\cite{Lyth:2001nq}) the full non-gaussian pdf is known and can be taken into account directly (e.g. Refs.~\cite{Young:2013oia,Bugaev:2013vba}).

Gamma-ray emission from Ultra Compact Mini Halos~\cite{Ricotti:2009bs,Scott:2009tu} (small dark matter halos which form at $z \sim 1000$ from smaller over-densities, with initial amplitude $\delta \sim 10^{-3}$) leads to tighter constraints on the primordial power spectrum than PBHs on scales $k \sim 1-10^{8} \, {\rm Mpc}^{-1}$~\cite{Josan:2010vn,Bringmann:2011ut}. However these constraints rely on the assumption that the dark matter is in the form of self-annihilating Weakly Interacting Massive Particles.

\subsection{Constraints on inflation models}
\label{sec-inf-const}

The amplitude of the primordial power spectrum on cosmological scales is measured to be ${\cal P}_{\cal R} (k \approx 10^{-3} \, {\rm Mpc}^{-1} ) \approx 10^{-10}$~\cite{Hlozek,Ade,Bird:2010mp} while the PBH limits are of order ${\cal P}_{\cal R}< 10^{-2}-10^{-1}$ on scales $k \sim 10^{-2}-10^{23} \, {\rm Mpc}^{-1}$. Therefore the PBH limits only constrain models in which the amplitude of the fluctuations is larger on small (physical) scales than on large scales (this is sometimes referred to as a `blue' spectrum).

In the mid-1990s it was found that for a power-law power spectrum, ${\cal P}_{\cal R} \propto k^{n_{\rm s}-1}$, where $n_{\rm s}$ is the spectral index, PBHs placed a constraint on the spectral index $n_{\rm s} < 1.25-1.30$~\cite{Carr:1993aq,Kim,Green:1997sz}. This was tighter than the CMB constraints at that time, however the spectral index is now accurately measured to be $n_{\rm s} = 0.9603 \pm 0.0073$  on cosmological scales~\cite{Ade}. Therefore if the power spectrum were a pure power law on all scales, then the number of PBHs formed would be completely negligible. However only very specific inflation potentials produce a constant spectral index (e.g. Ref. ~\cite{Vallinotto:2003vf}).
Generically the power spectrum will deviate from a pure power law  and, given the extremely wide range of scales probed by PBHs, it is possible for the amplitude of the perturbation to grow sufficiently with increasing wavenumber that PBHs could be produced in cosmologically interesting numbers. 

Refs.~\cite{Peiris:2008be,Josan:2010cj} used the flow equations for the evolution of the Hubble slow-roll parameters~\cite{Kinney:2992qn} to generate a large ensemble of inflation models. They found that a significant fraction of the models generated (in which inflation is terminated by an auxiliary field)  are compatible with all cosmological observations, but have perturbations on small scales which are sufficiently large to overproduce PBHs, and are hence excluded.

There are also specific inflation models that are constrained by PBH over-production.
In the running-mass inflation model the false vacuum dominated potential which arises in softly-broken global supersymmetry is flattened (so that slow-roll inflation, producing a close to scale-invariant power spectrum, can occur on cosmological scales) by quantum corrections~\cite{Stewart:1996ey,Stewart:1997wg}. However this can only happen over a limited range of scales, and over a large part of the parameter space the power spectrum grows significantly with decreasing scale and PBHs are over-produced~\cite{Leach:2000ea,Kohri:2007qn,Drees:2011hb}.
The power spectrum is also larger on small scales, potentially leading to PBH over-production~\cite{Kohri:2007qn}, in hill top inflation models~\cite{Boubekeur:2005zm}, where the potential flattens towards the end of inflation. In this case PBH constraints also exclude otherwise viable regions of parameter space~\cite{Alabidi:2009bk}. As we will discuss in Sec.~\ref{sec-dm}, PBHs can also be produced in interesting numbers in double or multiple-field inflation models, where the primordial perturbation spectrum has a spike on a particular scale.

In many inflation models the reheating process at the end of inflation starts with a period of parametric resonance known as preheating (see e.g. Ref.~\cite{Lyth:2009zz}  for an overview).  During preheating the amplification of field fluctuations can lead to the generation of large curvature perturbations and avoiding the over-production of PBHs constrains the couplings of the inflaton field~\cite{Green:2000he,Bassett:2000ha}.

\section{PBHs as dark matter}
\label{sec-dm}

Assuming general relativity is correct, there is extensive astronomical and cosmological evidence that the majority of the matter in the universe is in the form of non-baryonic cold dark matter (CDM). Since the CDM is non-baryonic, most candidates are new fundamental particles, for instance Weakly Interacting Massive Particles or axions. see e.g. Ref.~\cite{Bertone:2004pz}.
As PBHs form before nucleosynthesis they are non-baryonic. Therefore PBHs with $M_{\rm PBH}  >  10^{15} \, {\rm g}$, that have lifetime longer than the age of the Universe, are a potential CDM candidate. As we saw in Sec.~\ref{sec-abund} gamma-ray, lensing and dynamical constraints rule out PBHs with $10^{15}  \, {\rm g} < M_{\rm PBH} <10^{20} \, {\rm g}$ or $M_{\rm PBH}> 10^{25} \, {\rm g}$ making up all of the dark matter.  However this leaves a mass window ($10^{20} \, {\rm g} < M_{\rm PBH}< 10^{25} \, {\rm g}$) where PBHs can make up all of the CDM (see however Ref.~\cite{Pani:2014rca}). Unlike other CDM candidates, PBH are not a new fundamental particle. However, as we saw in Sec.~\ref{sec-form}, producing the large over-densities required for PBHs to form does require particle physics beyond the standard model.

Interest in PBHs as a DM candidate peaked in the late 1990s and early 2000s, due to the results at that time of microlensing searches towards the Large Magellanic Cloud. In their first two years of data the MACHO collaboration found 8 events,
significantly more than expected due to known stellar populations, and consistent with compact objects with $M_{\rm CO} \sim 0.5 M_{\odot} \approx 10^{33} \, {\rm g}$ making up roughly half of the mass of the MW halo~\cite{Alcock:1996yv}. With subsequent analysis of 5.7 years of data the best fit halo fraction dropped to $\sim 20\%$~\cite{Alcock:2000ph}.
Limits from star counts~\cite{Charlot} and chemical abundance constraints~\cite{Fields:1999ar} rule-out  baryonic objects, such as faint stars or white dwarves, comprising such a large fraction of the halo, and hence PBHs were an attractive MACHO candidate. 

PBHs with $M_{\rm PBH} \sim M_{\odot}$ would be formed at $t \sim 10^{-6} \, {\rm s}$, around the time of the QCD phase transition. 
If the QCD phase transition is first order the reduced pressure forces would lead to PBHs forming more easily at this epoch, however the amplitude of the primordial perturbations would still need to be significantly larger than on cosmological scales for an interesting number of PBHs to form~\cite{Jedamzik:1996mr,Schmid:1998mx,Jedamzik:1999am}. In other words, a feature in the primordial power spectrum is required. Such a feature can be produced in various ways. Ref.~\cite{Ivanov:1994pa} used a plateau in the inflation potential,  while Ref.~\cite{Yokoyama:1995ex} used multiple scalar fields to generate isocurvature perturbations and hence imprint a feature at a particular scale. In double inflation models, with either a single~\cite{Yokoyama:1998pt} or multiple scalar fields~\cite{Randall:1995dj,GarciaBellido:1996qt,Kawasaki:1997ju}, there are two periods of inflation and the perturbations on small scales, which are produced during the second period of inflation, can be significantly larger than those on cosmological scales.

As discussed in Sec.~\ref{sec-abund-lens-micro}, more recent microlensing results indicate that compact objects in the range $10^{25} \, {\rm g} < M_{\rm CO} < 10^{34} \, {\rm g}$ can not make up all of the DM in the MW halo~\cite{Tisserand:2006zx,Griest:2013aaa}. Lens in the Magellanic clouds and variable stars are now thought to account for some of the events found by the MACHO collaboration~\cite{Tisserand:2006zx}.
Interest in solar mass PBHs as a CDM candidate has therefore waned, however many of the models constructed to produce solar mass PBHs could, with different parameter values, also produce PBHs in the remaining allowed mass window (e.g. Refs.~\cite{Kanazawa:2000ea,Saito:2008em} for the case of double inflation).  Another possibility for producing PBH DM is an
inflaton potential with a step in its first derivative~\cite{Blais:2002nd}.
 
Ultimately the possibility of PBH CDM should be tested observationally. The constraints on PBHs with $M_{\rm PBH} \approx10^{25-26} \, {\rm g}$ come from a microlensing search using Kepler data~\cite{Griest:2013aaa}. Future data, from Kepler and WFIRST, will be sensitive to PBHs down to $M_{\rm PBH} \approx10^{24} \, {\rm g}$~\cite{Griest:2013aaa}.  
Future space-based gravitational wave detectors will be able to indirectly constrain PBH DM  produced from the collapse of large density perturbations in the mass range $10^{20} \, {\rm g} < M_{\rm PBH} < 10^{26} \, {\rm g}$ (as discussed in Sec. \ref{sec-abund-other-gw})~\cite{Saito:2009jt}. Future gravitational wave detectors could also detect PBHs with $10^{17}  \, {\rm g} < M_{\rm PBH} <10^{20} \, {\rm g}$ directly as they passes through the Solar System~\cite{Seto:2004zu}, while
PBHs with $M_{\rm PBH} \sim 10^{25} \, {\rm g}$ could be detected via pulsar timing using the Square Kilometer Array~\cite{Seto:2007kj}.
Finally the oscillations produced when a PBH with $M_{\rm PBH} > 10^{21} \, {\rm g}$ passes through a nearby star could be detected by the proposed Stellar Imager~\cite{Kesden:2011ij}.

\section{Summary}

In Sec.~\ref{sec-form} we reviewed the mechanisms via which PBHs can form, namely the collapse of large density fluctuations (Sec.~\ref{sec-form-dens}), cosmic string loops (Sec.~\ref{sec-form-cs}) or bubble collisions (Sec.~\ref{sec-form-bubble}). We concentrated mainly on the collapse of large density perturbations. After a review of the original pioneering calculations (Sec.~\ref{sec-form-dens-simp}) we described more recent developments (Sec.\ref{sec-form-dens-refine}), including the implications of critical collapse for the PBH mass function and determinations of the value of the threshold for PBH formation and its dependence on the shape of the density fluctuations.

We then looked at the observational constraints on the abundance of PBHs from the consequences of their evaporation products (Sec.~\ref{sec-abund-evap}), their lensing (Sec.~\ref{sec-abund-lens}) and dynamical effects (Sec.~\ref{sec-abund-dyn}) and their effects on other astrophysical objects and processes (Sec.~\ref{sec-abund-other}). The resulting limits on the fraction of the density of the Universe in the form of PBHs at the time they form, $\beta(M_{\rm PBH})$, depend on the PBH mass, $M_{\rm PBH}$, and lie in the range $10^{-5}$ to $10^{-30}$.

 In Sec.~\ref{sec-inf-translate} we explored how the PBH abundance constraints can be translated into limits on the primordial power spectrum, and how these limits can be used to constrain models of inflation (Sec.~\ref{sec-inf-const}). Avoiding PBH over-production constrains the parameter space of otherwise viable inflation models.
 
 Finally we discussed PBHs as a cold dark matter candidate (Sec.~\ref{sec-dm}). PBHs with mass in the range $10^{20} \, {\rm g} < M_{\rm PBH}< 10^{25} \, {\rm g}$ are a viable CDM candidate. To produce an interesting number of PBHs in this mass window requires a feature in the primordial power spectrum, and several inflation models have been constructed which achieve this. Various upcoming experiments/observations will be able to detect, or constrain, PBH DM.

%\begin{figure}[t]
%\sidecaption[t]
%\includegraphics[scale=.65]{figure}
%
% If no graphics program available, insert a blank space i.e. use
%\picplace{5cm}{2cm} % Give the correct figure height and width in cm
%\caption{If the width of the figure is less than 7.8 cm use the \texttt{sidecapion} command to flush the caption on the left side of the %page. If the figure is positioned at the top of the page, align the sidecaption with the top of the figure -- to achieve this you simply %need to use the optional argument \texttt{[t]} with the \texttt{sidecaption} command}
%\end{figure}

%
\begin{acknowledgement}
AMG is funded by the STFC.
\end{acknowledgement}
%
%\section*{Appendix}
\addcontentsline{toc}{section}{Appendix}
\input{referenc-green}
\end{document}

%% file: referenc-green.tex
\bigskip

% Use the following (APS) syntax and markup for your references if 
% the subject of your book is from the field 
% "Mathematics, Physics, Statistics, Computer Science"
%
% Online Document
%\bibitem{phys-online} J. Dod, in \textit{The Dictionary of Substances and Their Effects}, Royal Society of Chemistry. (Available via DIALOG, 1999), 
%\url{http://www.rsc.org/dose/title of subordinate document. Cited 15 Jan 1999}
%
% Monograph
%\bibitem{phys-mono} H. Ibach, H. L\"uth, \textit{Solid-State Physics}, 2nd edn. (Springer, New York, 1996), pp. 45-56 
%
% Journal article
%\bibitem{phys-journal} S. Preuss, A. Demchuk Jr., M. Stuke, Appl. Phys. A \textbf{61}
%
% Journal article by DOI
%\bibitem{phys-DOI} M.K. Slifka, J.L. Whitton, J. Mol. Med., doi: 10.1007/s001090000086
%
% Contribution 
%\bibitem{phys-contrib} S.E. Smith, in \textit{Neuromuscular Junction}, ed. by E. Zaimis. Handbook of Experimental Pharmacology, vol 42 (Springer, Heidelberg, 1976), p. 593
%
%